\documentclass[11pt]{article}
\usepackage{amssymb,amsmath}
\usepackage{graphicx}
\usepackage{amsmath}
\usepackage{amsfonts}
\usepackage{amssymb}%
\newtheorem{theorem}{Theorem}
\newtheorem{criterion}[theorem]{Criterion}

\begin{document}

\title{How to Formulate Non-Equilibrium \\Local States in QFT?\\-- General Characterization and \\\ \ Extension to Curved Spacetime--\\\medskip{\normalsize \textsl{Dedicated to Professor Hiroshi Ezawa}}\\{\normalsize \textsl{on the occasion of his seventieth birthday}}}
\author{Izumi OJIMA\\Research Institute for Mathematical Sciences, \\Kyoto University, Kyoto 606-8502}
\maketitle

\begin{abstract}
The essence of a general formulation to accommodate non-equilibrium local
states in relativistic quantum field theory is explained from the viewpoint of
comparison at a spacetime point between unknown generic states to be
characterized as such states and the known family of probabilistic mixtures of
equilibrium states. Taking advantage of the local nature of the problem, we
extend the formalism to the general-relativistic context with curved spacetimes.

\end{abstract}

\section{Introduction}

It is a great honour for me to make a contribution to this volume to
commemorate Professor Ezawa's 70th anniversary, especially because I have
benefited very much from him and from what he wrote and said on physics in
general and on quantum field theory in particular, at various stages of my
research career. I have been impressed by his wide perspectives ranging from
mathematical physics to any kind of physical aspects of nature, among which
non-equilibrium statistical physics to be discussed in the following, is an
important common subject between him and myself.

\textquotedblleft Non-equilibrium\textquotedblright\ seems to be one of the
characteristic features of domains or phenomena in which nature exhibits its
most vivid essence. However, theoretical attempts of systematic approaches to
it starting\textit{\ from the first principles of microscopic quantum theory}
seem to have been rather rare (aside from some attractive phenomenological
theories in thermodynamic frameworks), in sharp contrast with equilibrium
cases. For the latter, we know the existence of variety of achievements
successfully attained in the clear-cut formulation\textit{\ }based upon the
notion of Gibbs ensembles or \textit{Kubo-Martin-Schwinger (KMS) states} (as
generalized version of the former applicable to infinitely extended systems in
thermodynamic limit), ranging from detailed analyses of concrete models to
abstract sophisticated mathematical treatments of general infinite systems.
The reason for such a difference seems to be evident: pursuits for concise and
universal characterization of non-equilibrium have been given up, for such
reasons as

\begin{itemize}
\item[i)] strong negative influence of \textit{poor and ambiguous images}
originating from negative\textit{\ }ideas and pictures: in the word
\textquotedblleft\textit{non}-equilibrium (states)\textquotedblright\ one sees
only \textit{simple negation} of equilibrium, missing positive contents.

\item[ii)] On the positive side, the experiences of being confronted with the
huge \textit{variety} exhibited by non-equilibrium domains can easily mislead
one to a superstition that emphasis on \textit{ample individual features at
macroscopic levels} is equivalent to negating connections with their universal
microscopic bases, which ends up with pessimism against \textit{ab initio}
discussions starting from the \textquotedblleft first
principles\textquotedblright\ of microscopic quantum theory.
\end{itemize}

We need to recall, however, that the great achievements in equilibrium
statistical mechanics and solid state physics should be found in their
\textit{unified understanding of macroscopic variety on the universal basis of
microscopic quantum theory}. At this point, we recall also many fruitful
positive examples in the history of transitions to such domains with
\textquotedblleft non-\textquotedblright, as $\langle\langle$from Euclidean to
\textit{non}-Euclidean geometries$\rangle\rangle$, $\langle\langle$from
commutative classical world to \textit{non-}commutative quantum one$\rangle
\rangle$, or $\langle\langle$from standard to \textit{non}-standard
logic$\rangle\rangle$, and so on.

Taking analogy to the basic idea of \textit{manifolds }exhibiting the process
$\langle\langle$from Euclidean to \textit{non}-Euclidean geometries$\rangle
\rangle$, I try here to re-view the general conceptual essence of our recent
work \cite{BOR, Oji02} towards a general framework for treating nonequilibrium
in relativistic QFT, where nonequilibrium local states are specified by a
concise \textit{selection criterion} and their thermal
\textit{interpretations} are canonically fixed. (The emphasis here is on the
conceptual aspects with technical details omitted.)

When we try to understand non-Euclidean geometry as the geometry of curved
spaces $M$, e.g., surface of the earth, in a precise way, the aim is never
attained with such insistence that what is curved should be treated as it
stands. According to the common wisdom, we start from a small neighbourhood
$U_{i}$ in $M$ so that effects of curvature are negligible and try to draw a
precise \textit{map} of $U_{i}$ \textit{on a flat Euclidean space}, which is
nothing but a usual local map in the case of surface of the earth. While the
whole sphere cannot correctly be drawn on one sheet, a (geometrically)
\textit{\textquotedblleft precise\textquotedblright} description of such a
curved space as the earth sphere can be attained by an \textit{atlas} as the
totality of many local charts $\varphi_{i}:U_{i}\rightarrow\mathbb{R}^{d}$
covering $M=\cup_{i}U_{i}$.

This familiar discussion found at the beginning of any textbooks on manifolds
tells us the following points: In our scientific attempts to describe
something in nature, e.g., \textquotedblleft curved space\textquotedblright%
\ $M$, as

\begin{itemize}
\item[i)] our \textit{unknown target object} to be described,
\end{itemize}

\noindent we inevitably need to relate it with

\begin{itemize}
\item[ii)] something familiar\ to serve as a\textit{\ standard reference
frame},
\end{itemize}

\noindent(such as a flat Euclidean space $\mathbb{R}^{d}$ in the example),
which is implemented by

\begin{itemize}
\item[iii)] the processes of measuring unknown target objects so\ that
\textit{i) is mapped to ii)} [: as the case of $\varphi_{i}$ above],
\end{itemize}

\noindent and then,

\begin{itemize}
\item[iv)] the data obtained in iii) need be collected and \textit{organized
into a coherent interpretation} (through which a description is realized) [:
the atlas in the example].
\end{itemize}

These are just the minimum ingredients for our purpose. What corresponds to
these in our discussion of non-equilibrium local states can be identified as follows:

\begin{itemize}
\item[i')] [\textit{unknown target object}]=unknown quantum state $\omega$ to
be identified as our non-equilibrium local state,

\item[ii')] [\textit{standard reference frame}]=[\textit{family }$K$
\textit{of thermal reference states}]\textit{\ }consisting of convex
combinations of global equilibrium states $\omega_{\beta}$ at all the possible temperatures,

\item[iii')] [processes of measuring to \textit{map i') to ii')}%
]=[\textit{local thermal observables }$\mathcal{T}$] which detect local
thermal properties and whose measured values in an unknown state $\omega$ in
i') are to be compared with the corresponding values in known states belonging
to $K$,

\item[iv')] [\textit{to organize data obtained in iii') into a coherent
interpretation}]= [\textit{criterion to select non-equilibrium local states} +
\textit{thermal interpretations} of selected states] in terms of measured
values of thermal quantities obtained in iii').
\end{itemize}

\noindent In the following sections, the actual contents of ii')-iv') will be explained.

\section{Thermal reference states}

In relativistic QFT we identify global thermal equilibria with relativisitc
\textbf{KMS states} given as follows. The KMS condition \cite{HaHuWi,BrRo} is
a mathematical characterization of a KMS state $\omega_{\beta}$ which
generalizes familiar Gibbs states into a form applicable to infinitely
extended systems by extracting characteristic relation
\begin{equation}
Tr(e^{-\beta H}AB(t))=Tr(e^{-\beta H}Ae^{iHt}Be^{-iHt})=Tr(e^{-\beta
H}B(t-i\beta)A).
\end{equation}
In a relativistic version, a state\ $\omega_{\beta}$ as an \textit{expectation
functional} on the algebra $\mathcal{A}$ of observables ($\omega_{\beta
}:\mathcal{A}\ni A\longmapsto\omega_{\beta}(A)\in\mathbb{C}$) is called a
\textit{relativistic KMS state} with an inverse temperature 4-vector
$\beta=(\beta^{\mu})\in V_{+}(:=\{x\in R^{4};x^{2}\equiv(x^{0})^{2}-(\vec
{x})^{2}>0,x^{0}>0\})\,$, $\,$if it satisfies the following
\textit{relativistic KMS condition} \cite{BrBu}: for any pair $A,B\in
\mathcal{A}$ there is a function $h=h_{A,B}$, analytic in $D_{\beta
}:=\mathbb{R}^{4}+i\,\left(  V_{+}\cap(\beta-V_{+})\right)  $, and continuous
on $\overline{D_{\beta}}$ with the boundary conditions
\begin{equation}
h(a)=\omega_{\beta}(A\alpha_{a}(B)),\ \ h(a+i\beta)=\omega_{\beta}(\alpha
_{a}(B)A),
\end{equation}
where $\mathbb{R}^{4}\ni a\longmapsto\alpha_{a}\in Aut(\mathcal{A})$ [:
*-automorphism group of $\mathcal{A}$] is a spacetime translation acting on
$\mathcal{A}$. Then, $\omega_{\beta}$ can be seen to describe a global thermal
equilibrium at a temperature $T=(k_{B}\sqrt{\beta^{2}})^{-1}$ in a
\textit{rest frame} determined by a timelike unit vector $e=\beta/\sqrt
{\beta^{2}}\in V_{+}$.

The totality $K_{\beta}$ of relativistic KMS state with $\beta\in V_{+}$ is
known \cite{BrRo} to be a \textit{simplex} admitting for each state $\in$
$K_{\beta}$ a unique decomposition into a convex combination of extremal
points, which can be identified with \textit{thermodynamic pure phases} and
can be parametrized uniquely by such thermodynamic parameters as $\beta$ (in
combination with some such additional ones as chemical potentials $\mu$, if necessary).

A non-trivial Poincar\'{e}\thinspace\ transformation $\lambda\!=\!(a,\Lambda
)\in\mathcal{P}_{+}^{\uparrow}:=\mathbb{R}^{4}\rtimes L_{+}^{\uparrow}$
transforms $\omega_{\beta}\in K_{\beta}\,$ into another one $\omega_{\beta
}\circ\alpha_{\lambda}^{-1}\in K_{\Lambda\beta}$ with inverse temperature
$\Lambda\beta$, through which a temparature defined by $1/\sqrt{\beta^{2}%
}=1/\sqrt{(\Lambda\beta)^{2}}$ is unchanged but the state $\omega_{\beta}%
\circ\alpha_{\lambda}^{-1}\neq\omega_{\beta}$ can be different from the
original one $\omega_{\beta}$, because of the change of reference frame
$e=\beta^{\mu}/\sqrt{\beta^{2}}\rightarrow\Lambda e\neq e$: this is the
spontaneous breakdown of Lorentz invariance due to temperature \cite{Oj}. When
convenient for simplification in the following, we introduce the assumption of
the \textit{absence of phase transitions} formulated as the uniqueness of KMS
state at each temperature $\beta\in V_{+}$, which implies the relation
$\omega_{\beta}\circ\alpha_{\lambda}^{-1}=\omega_{\Lambda\beta}$, and hence,
the invariance of $\omega_{\beta}$ under spacetime translations as well as its
isotropy in the rest frame.

As ii') [\textit{family }$K$ \textit{of thermal reference states}] for local
thermal description of an unknown state $\omega$, we take all the possible
statistical mixtures $\omega=\int_{B}d\rho(\beta)\,\omega_{\beta}%
=:\omega_{\rho}$ of KMS states $\omega_{\beta}$ in which temperature $\beta$
is fluctuating over some compact subsets $B$ in $V_{+}$ with such a
probability distribution $d\rho(\beta)$ that supp$(\rho)\subset B$. Namely, we
adopt the definition
\begin{equation}
K:=\underset{B:\text{ cpt }\subset V_{+}}{\bigcup}\,K_{B}%
\end{equation}
with
\begin{equation}
K_{B}:=\{\omega_{\rho}=\int_{B}d\rho(\beta)\,\omega_{\beta};\ \rho\text{:
probability measure on }V_{+}\text{ with supp}(\rho)\subset B\}.
\end{equation}
In more general situations, the requirement of compact supports of $\rho$ may
have to be removed and, when the assumption of no phase transition is
invalidated, we also add other order-parameters like chemical potential $\mu$,
etc., so as for all the relevant thermodynamic pure phases to be
discriminated. Then we denote generically the spaces of all the relevant
thermodynamic parameters $(\beta,\mu)$ and of (a suitable class of)
probability measures $d\rho(\beta,\mu)$, respectively, by $B_{K}$ and by $Th$
($=M_{1}(B_{K})=$ the space of probability measures on $B_{K}$, for instance):
$K\ni\omega_{\rho}:=$ $\int_{B_{K}}d\rho(\beta,\mu)\,\omega_{\beta,\mu}$,
$(\beta,\mu)\in B_{K}$, $\rho\in Th$.

\section{\textquotedblleft Coordinatization\textquotedblright\ by local
thermal observables}

Next, we need \textquotedblleft coordinatization map\textquotedblright\ iii')
connecting i') to ii'). The basic idea for this is first to measure some
thermal observables $\{\Phi_{i}\}$ in an unknown state $\omega$ under
consideration and then to plot on \textquotedblleft$\{\Phi_{i}\}$%
-space\textquotedblright\ the measured values $\{\omega(\Phi_{i})=\Phi
_{i}(\omega)\}$ regarding them as \textquotedblleft$\Phi_{i}$%
-coordinates\textquotedblright\ of $\omega$. If we find a data set $\{\Phi
_{i}(\omega_{\beta})\}$ for a known equilibrium state $\omega_{\beta}\in
K_{\beta}$ s.t. $\Phi_{i}(\omega)=\Phi_{i}(\omega_{\beta})$, then our unknown
$\omega$ can be identified with this known $\omega_{\beta}$ $\in K $
\textit{as far as thermal properties determined by quantities }$\{\Phi_{i}%
\}$\textit{\ is concerned}: $\omega\underset{\{\Phi_{i}\}}{\equiv}%
\omega_{\beta}$. (The inclusion of mixture $\omega_{\rho}=\int_{B}d\rho
(\beta)\,\omega_{\beta}$ is just for the sake of wider range of data search to
include temperature fluctuations.)

Now the problem is \textit{how to find physical quantities} $\{\Phi_{i}%
\}$\textit{\ suitable for describing local thermal properties of
non-equilibrium states in the framework of relativistic QFT?}

The aim of this section is to give an answer to this. While the most desirable
form of the answer would be such that thermal properties of unknown states in
a small spacetime region $\mathcal{O}$ are determined by local observables
measurable within $\mathcal{O}$. In contrast to our motivating example of
manifold, however, it is almost impossible to attain directly this goal
starting from a finitely extended region.

Our strategy here is \textit{first to concentrate on a spacetime point} $x$
and \textit{then to extend the obtained results to a finitely extended
region}. However, we immediately encounter the well-known difficulty of
ultraviolet (UV) divergences invalidating a quantum field at a point, which
seems to make our desire hopeless! (Perhaps, this may be one of the reasons
for which the programme to construct general framework of non-equilibrium
starting from microscopic quantum theory has been discouraged for a long
time.) It is too early, however, to give up here! \textit{There is an escape
through which quantum field }$\hat{\phi}(x)$\textit{\ at a point can be made
meaningful.} Since the cause to invalidate $\hat{\phi}(x)$ is just the UV
divergenges due to high-frequency modes in quantum fields, $\hat{\phi}%
(x)$\textit{\ makes sense in states to which high energy components make no
significant contributions}.

This idea can be formulated in a mathematically meaningful form as follows. In
many model examples in constructive field theory, the validity of
\textit{energy-bound inequality} has been checked \cite{FrHe}: for $\forall
l>0$ there exist $m>0$ and a constant $c>0$ s.t.
\begin{equation}
||(\mathbf{1}+H)^{-m}\,\hat{\phi}(f)\,(\mathbf{1}+H)^{-m}||\leq c\,\int
\!dx\,|(\mathbf{1}-\Delta)^{-l}f(x)|,
\end{equation}
holds for $\forall f\in\mathcal{S}(\mathbb{R}^{4})$. Here $H$ is a Hamiltonian
defined in the vacuum representation where operator norm $||\cdot||$ is
defined and $\Delta$ is the Laplacian in $\mathbb{R}^{4}$. Taking a sequence
$\delta_{i}$ of test functions s.t. $\delta_{i}\underset{i\mathbb{\rightarrow
}\infty}{\rightarrow}\delta_{x}$ (: Dirac measure on $x$), we have, for
sufficiently large $m>0$,
\begin{equation}
\lim_{i\rightarrow\infty}\,(\mathbf{1}+H)^{-m}\,\hat{\phi}(\delta
_{i})\,(\mathbf{1}+H)^{-m}=:(\mathbf{1}+H)^{-m}\,\hat{\phi}(x)\,(\mathbf{1}%
+H)^{-m}\
\end{equation}
which justifies $\hat{\phi}(x)$ mathematically. Then, $\omega(\hat{\phi}(x)) $
is meaningful for any state $\omega$ s.t. $\omega((\mathbf{1}+H)^{2m})<\infty$.

By replacing $H$ meaningful only in the vacuum represetation with a
\textit{local Hamiltonian} $H_{\mathcal{O}}$ playing the role of $H$ in a
local region $\mathcal{O}$ \textit{independently of representations}, we
arrive at the condition
\begin{equation}
\omega((\mathbf{1}+H_{\mathcal{O}})^{2m})<\infty, \label{regularity}%
\end{equation}
to be imposed on states $\omega$ in question. Actually this is automatically
satisfied by any states admitting local thermal interpretation which should
have finite energy locally. We denote $E_{\mathcal{O}}$ the totally of states
$\omega$ satisfying Eq.(\ref{regularity}) with a suitable $m>0$,
\begin{equation}
E_{\mathcal{O}}:=\{\omega;\omega\text{: state of }\mathcal{A}\text{ and
}\exists m>0\text{ s.t. }\omega((\mathbf{1}+H_{\mathcal{O}})^{2m}%
)<\infty\}\supset K,
\end{equation}
whose pointlike limit (projective limit)
\begin{equation}
E_{x}(=\underset{\mathcal{O}\rightarrow x}{\underleftarrow{\lim}%
}E_{\mathcal{O}})\hookleftarrow K \label{germ}%
\end{equation}
is given by the set of equivalence classes in $\cup_{\mathcal{O}%
}E_{\mathcal{O}}$ with respect to the equivalence relation $\thicksim
$\ defined by
\begin{equation}
\omega_{1}\thicksim\omega_{2}\overset{\mathrm{def}}{\Longleftrightarrow
}\exists\mathcal{O}\text{: neighbourhood of }x\text{ s.t. }\omega
_{1}\upharpoonright_{\mathcal{O}}=\omega_{2}\upharpoonright_{\mathcal{O}}.
\end{equation}
(The family $\mathcal{O}\longmapsto E_{\mathcal{O}}$ constitutes a presheaf of
state germs \cite{HO} whose stalk at $x$ is given by $E_{x}$.)

While the product structure of quantum fields is lost through this procedure,
it can effectively be recovered by the notion of \textit{normal products} in
the operator-product expansion (OPE) reformulated recently by \cite{Bo} in a
mathematically rigorous form. Namely, \thinspace linear spaces $\mathcal{N}%
(\hat{\phi}^{2})_{\,q,x}$ consisting of normal products appearing in the
expansion of $\hat{\phi}(x+\zeta)\hat{\phi}(x-\zeta)$ around $\zeta=0$ (valid
for sufficiently large $n\in\mathbb{N}$),
\begin{equation}
||(\mathbf{1}+H_{\mathcal{O}})^{-n}\left[  \hat{\phi}(x+\zeta)\hat{\phi
}(x-\zeta)-\sum_{j=1}^{J(q)}c_{j}(\zeta)\,\hat{\Phi}_{j}(x)\right]
(\mathbf{1}+H_{\mathcal{O}})^{-n}||\leq c^{\prime}\,|\zeta|^{q},
\label{wilson}%
\end{equation}
as coefficients $\hat{\Phi}_{j}(x)$ of $c$-number singular functions
$c_{j}(\zeta)$ in $\zeta$ are seen to serve as substitutes for the ill-defined
$\hat{\phi}(x)^{2}$, and similarly $\mathcal{N}(\hat{\phi}^{p})_{\,q,x}$ for
higher power $\hat{\phi}(x)^{p}$.

Through the similar expansion of $\partial_{\zeta}\,\hat{\phi}(x+\zeta
)\hat{\phi}(x-\zeta)$, the derivatives in the relative coordinates $\zeta$
(called \textquotedblleft\textit{balanced derivatives}\textquotedblright%
\ here),
\begin{equation}
||(\mathbf{1}+H_{\mathcal{O}})^{-n}\left[  \partial_{\zeta}\,\hat{\phi
}(x+\zeta)\hat{\phi}(x-\zeta)-\sum_{j=1}^{J(q)}\partial_{\zeta}\,c_{j}%
(\zeta)\ \hat{\Phi}_{j}(x)\right]  (\mathbf{1}+H_{\mathcal{O}})^{-n}||\leq
c^{\prime\prime}\,|\zeta|^{r}, \label{balanced}%
\end{equation}
can be similarly made meaningful (for sufficiently large $n,q\in\mathbb{N}$),
which describe internal structures of composite operators at the same
spacetime point $x$. The expectation values of these normal products determine
$p$-point correlation functions around $x$. While derivatives $\partial_{x}$
in the centre of mass coordinates are sensitive to the spacetime inhomogeneity
of an unknown state $\omega$, the corresponding quantities to $\omega
(\partial_{x}(\cdots))$ in thermal reference states are all vanishing
$\omega_{\beta}(\partial_{x}(\cdots))=0$, owing to the translational
invariance of $\omega_{\beta}$. Since the comparison iii') between $\omega$
and $\omega_{\beta}$ is for the sake of clarifying thermal properties of
$\omega$ instead of spacetime ones, this discrepancy indicates that local
observables involving $\partial_{x}$ should not be counted as \textit{local
thermal observables} suitable for detecting local thermal properties. With all
such irrelevant observables excluded, a suitable choice of \textit{local
thermal observables} as \textquotedblleft coordinatization
map\textquotedblright\ in iii') amounts to the \textit{linear space
}$\mathcal{T}_{x}$\textit{\ of point-like fields,}
\begin{equation}
\mathcal{T}_{x}\,:=\,\sum_{p,q}\,\mathcal{N}(\hat{\phi}_{0}^{\,p})_{\,q,x}\,,
\end{equation}
consisting of basic fields $\hat{\phi}_{0}(x)$ at $x$ together with their
normal products $\mathcal{N}(\hat{\phi}_{0}^{p})_{\,q,x}$.\ 

What is remarkable about $\mathcal{T}_{x}$ is its natural \textit{hierarchical
nesting structure} ordered by indices $m,p,q$ related to energy bound and OPE,
according to their increasing orders starting from scalar multiples of
identity with basic fields $\hat{\phi}_{0}(x)$ coming next, and so on. Since
$p$-point functions with larger $p$ govern those with smaller $p $, this
hierarchy has an operationally intrinsic meaning in such a form as
\textquotedblleft the larger $p$, the \textit{finer resolution of thermal
properties }is provided by $\mathcal{N}(\hat{\phi}_{0}^{\,p})_{\,q,x}$,
$q>0$\textquotedblright. In this context, macroscopic properties of thermal
states are expected to be described by subspaces $\mathcal{N}(\hat{\phi}%
_{0}^{\,p})_{\,q,x}$\thinspace\ with smaller $p$,$q$.

\noindent\newline---Macroscopic interpretations of $\mathcal{T}_{x}$---

We now examine how these local thermal observables in $\mathcal{T}_{x}$
provide information about macroscopic thermal properties of states in $K$.
This is materialized by \noindent\textit{thermal functions as macroscopic
observables }as follows. In thermodynamics, the physical contents of relevant
thermodynamic quantities like internal energy, entropy, etc., are specified by
their dependence on temperature (together with other necessary thermodynamic
parameters like pressure, chemical potentials, etc.). In parallel with this,
all intensive thermal parameters associated with states in $K$ can be
represented here by functions $B_{K}\ni(\beta,\mu)\mapsto F(\beta,\mu)$, which
we call \textit{thermal functions. }

The relation between \textit{quantum local thermal observables} and
\textit{classical macroscopic observables} is described by a map $\mathcal{C}$
associating a thermal function $\mathcal{C}(A)$ to each quantum observable
$A\in\mathcal{A}$ or $\mathcal{T}_{x}$ by%
\begin{equation}
\mathcal{C}:A\longmapsto\mathcal{C}(A):=[(\beta,\mu)\longmapsto\omega
_{\beta,\mu}(A)]\in C(B_{K}).
\end{equation}
In the case where all the thermal reference states $\omega_{\beta}$ are
translation invariant, the thermal function $\mathcal{C}(\hat{\Phi}%
(x))=\Phi_{x}$ corresponding to $\hat{\Phi}(x)\in\mathcal{T}_{x}$ is
$x$-independent, which is invalidated, for instance, by the crystalline
structures to break the spatial homogeneity, though. In the special case of
\textit{no }phase transitions where we have $\omega_{\beta}\circ
\alpha_{(\Lambda,a)}^{-1}=\omega_{\Lambda\beta}$, the thermal function
$\mathcal{C}(\hat{\Phi}(x))=\Phi$ even becomes an $x$-independent Lorentz
tensor in $\beta$.

We see now that thermal interpretation of each $\hat{\Phi}(x)\in
\mathcal{T}_{x}$ is given by thermal function $(\beta,\mu)\longmapsto\Phi
_{x}(\beta,\mu)=\mathcal{C}(\hat{\Phi}(x))(\beta,\mu)=\omega_{\beta,\mu}%
(\hat{\Phi}(x))$ (which amounts to recording the mean values of a local
thermal observable $\hat{\Phi}(x)${\small \ }in all equilibrium
states{\small \ }$\omega_{\beta,\mu}$).

Since the map $\mathcal{C}$ is normalized and positive linear, $\mathcal{C}%
(\mathbf{1})=1,\mathcal{C}(\hat{A}^{\ast}\hat{A})\geq0$ taking values in a
commutative algebra $C(B_{K})$,\thinspace\ it is a \textit{completely
positive} (CP) map characterized by the condition $\sum_{ij=1}^{n}\bar{f}%
_{i}\mathcal{C}(\hat{A}_{i}^{\ast}\hat{A}_{j})f_{j}\geq0$ for $\forall
n\in\mathbb{N},\forall f_{1},\cdots,\forall f_{n}\in C(B_{K})$ and
$\forall\hat{A}_{1},\cdots,\forall A_{n}\in\mathcal{A}$. The dual map
$\mathcal{C}^{\ast}$ of CP map $\mathcal{C}$ defined on states by%

\begin{align}
&  \mathcal{C}^{\ast}(\rho)(\hat{A})=\rho(\mathcal{C}(\hat{A}))=\int_{B_{K}%
}d\rho(\beta,\mu)\mathcal{C}(\hat{A})(\beta,\mu)=\int_{B_{K}}d\rho(\beta
,\mu)\omega_{\beta,\mu}(\hat{A}),\nonumber\\
&  \Longrightarrow\mathcal{C}^{\ast}(\rho)=\int_{B_{K}}d\rho(\beta,\mu
)\omega_{\beta,\mu}=\omega_{\rho}\in K,
\end{align}
becomes a \textit{\textbf{c}lassical-\textbf{q}uantum (c}$\rightarrow
$\textit{q) channel} \cite{OhyaPetz} $\mathcal{C}^{\ast}:Th\ni\rho
\longmapsto\mathcal{C}^{\ast}(\rho)\in K$, mapping classical probabilities
$\rho$ into quantum states $\mathcal{C}^{\ast}(\rho)\in K$. (Recall that
$Th=M_{1}(B_{K})$ is the space of classical thermal states identified with
probability measures $\rho$ on $B_{K}$ describing the mean values of
thermodynamic parameters $(\beta,\mu)$ together with their fluctuations.)
Measuring a local thermal observable $\hat{\Phi}(x)\in\mathcal{T}_{x}$ in this
thermal reference state $\mathcal{C}^{\ast}(\rho)$, we obtain
\begin{equation}
\mathcal{C}^{\ast}(\rho)(\hat{\Phi}(x))=\int_{B_{K}}\!d\rho(\beta,\mu
)\,\omega_{\beta,\mu}(\hat{\Phi}(x))=\int_{B_{K}}\!d\rho(\beta,\mu
)[\,\mathcal{C}(\hat{\Phi}(x))](\beta,\mu)=\rho(\Phi_{x}).
\end{equation}
Thus the \textit{thermal interpretation} of a quantum observable $\hat{\Phi
}(x)$\ in all thermal reference states of the form $\mathcal{C}^{\ast}%
(\rho)=\omega_{\rho}\in K$ is given by the corresponding macroscopic
\textit{thermal function} $\mathcal{C}(\hat{\Phi}(x))$ evaluated with the
classical probability\textit{\ }$d\rho(\beta,\mu)$ which describes the
\textit{fluctuations} of thermodynamic configurations $(\beta,\mu)$ in
$\omega_{\rho}$.

This applies to the case where $\rho$ is already known. What we need in the
actual situations is how to determine the \textit{unknown} $\rho$ from the
given data list $\Phi\longmapsto\omega_{\rho}(\hat{\Phi}(x))=\rho(\Phi_{x})$
of expectation values of thermal functions $\Phi_{x}=\mathcal{C}(\hat{\Phi
}(x))$ (which is the problem of state estimation): this problem can be solved
if $\mathcal{T}_{x}$ has sufficiently many local thermal observables so that
the image $\mathcal{C}(\mathcal{T}_{x})$ of $\mathcal{T}_{x}$ is dense in
$C(B_{K})$ so as to approximate arbitrary continuous functions of $(\beta
,\mu)$ (which need be checked in each concrete model). In this case $\rho$ is
given as the unique solution to a (generalized) \textquotedblleft moment
problem\textquotedblright. Thus we see:

\begin{itemize}
\item[$\bigstar$] If the set $\mathcal{T}_{x}$\textit{\ }of local thermal
observables is large enough to discriminate\ all the thermal reference states
in $K$, any reference state $\in K$ can be written as $\mathcal{C}^{\ast}%
(\rho)$ in terms of a uniquely determined probability measure $\rho$ on
$B_{K}$ describing the statistical fluctuations of thermal parameters in the
state in question. Then local thermal observables\textit{\ }$\hat{\Phi}%
(x)\in\mathcal{T}_{x}$\textit{\ }provide the same information on the thermal
properties of states in $K$ as that provided by the corresponding classical
macroscopic thermal functions\textit{\ }$\Phi=\mathcal{C}(\hat{\Phi})$\ [e.g.,
internal energy, entropy density, etc.]: $\omega_{\rho}(\hat{\Phi})=\rho
(\Phi)$.\newline
\end{itemize}

In this situation, \textit{any continuous function} $F$ on compact $B\subset
V_{+}$ can be approximated by thermal functions $\Phi_{x}=\mathcal{C}%
(\hat{\Phi}(x))$ with arbitrary precision, \textit{even if} $F$ itself is
\textit{not} an image of $\mathcal{C}$. In spite of the absence of quantum
$\hat{s}(x)\in\mathcal{T}_{x}$ s.t. $\omega_{\beta}(\hat{s}(x))=s(\beta)$:
\textit{entropy density}, $s(\beta)$ can be treated as an \textit{approximate}
thermal function.

The above ($\bigstar$) ensures the existence of \textit{inverse of
c}$\rightarrow$\textit{q channel} $\mathcal{C}^{\ast}$ on $K$:
\begin{equation}
K\ni\omega_{\rho}=\mathcal{C}^{\ast}(\rho)\longleftrightarrow(\mathcal{C}%
^{\ast})^{-1}(\omega_{\rho})=\rho\in Th,
\end{equation}
and the thermal interpretation of thermal reference states $\in K$ is just
given by this \textbf{\textit{\textbf{q} }}$\rightarrow$\textit{\textbf{c
channel }}$(\mathcal{C}^{\ast})^{-1}:K\ni\omega\longmapsto\rho\in Th$ s.t.
$\omega=\mathcal{C}^{\ast}(\rho)$ \cite{Oji02}(, which can be regarded as a
simple adaptation and extension of the notions of classifying spaces and
classifying maps to the context involving (quantum) probability theory). We
express formally the essence of the above situation ($\bigstar$) by
\begin{equation}
K(\omega,\mathcal{C}^{\ast}(\rho))/\mathcal{T}_{x}\overset{q\rightleftarrows
c}{\simeq}Th((\mathcal{C}^{\ast})^{-1}(\omega),\rho)/\mathcal{C}%
(\mathcal{T}_{x}), \label{adjunction1}%
\end{equation}
with a quantum state $\omega\in E_{x}$ and a probability measure $\rho\in Th$.
(The precise meaning of this can be understood as a categorical adjunction
between two functors given by \textit{c }$\rightarrow$\textit{q}
($\mathcal{C}^{\ast}$) and \textit{q }$\rightarrow$\textit{c }($(\mathcal{C}%
^{\ast})^{-1}$) channels which connect $K$ and $Th$ both regarded as
groupoids\footnote{A groupoid $\Gamma$ is, roughly speaking, a generalization
of a group so that there are many unit elements constituting a set $\Gamma
_{0}$. Each element $\gamma\in\Gamma$ has its source $s(\gamma)$ and target
$r(\gamma)$ in $\Gamma_{0}$ and these points are thought to be connected by
$\gamma$, $s(\gamma)\overset{\gamma}{\rightarrow}r(\gamma)$, in an invertible
way: $r(\gamma)\overset{\gamma^{-1}}{\rightarrow}s(\gamma)$. Two elements
$\gamma_{1},\gamma_{2}\in\Gamma$ are not always composable but $\gamma
_{1}\gamma_{2}$ is meaningful only when $r(\gamma_{2})=s(\gamma_{1})$. There
is a one-to-one and onto correspondence between a groupoid $\Gamma$ and an
\textit{equivalence relation} $\thicksim$\ on a set $\Gamma_{0}$ through
[$a\thicksim b$ for $a,b\in$ $\Gamma_{0}$] $\Longleftrightarrow$
[$\exists\gamma\in\Gamma$ s.t. $a=r(\gamma)$ and $b=s(\gamma)$]. As a
category, $\Gamma$ is one with $\Gamma_{0}$ as the set of objects and with all
its morphisms being invertible. In our case, they are defined by $\Gamma
_{0}:=K$ or $Th$ together with the equivalence relation $\underset
{\mathcal{T}_{x}}{\equiv}$ or $\underset{\mathcal{C}(\mathcal{T}_{x})}{\equiv
}$, respectively.} corresponding to the equivalence relations $\omega
_{1}\underset{\mathcal{T}_{x}}{\equiv}\omega_{2}$ and $\rho_{1}\underset
{\mathcal{C}(\mathcal{T}_{x})}{\equiv}\rho_{2}$ defined, respectively, by
$(\omega_{1}-\omega_{2})(\mathcal{T}_{x})=\{0\}$ and $(\rho_{1}-\rho
_{2})(\mathcal{C}(\mathcal{T}_{x}))=\{0\}$; in the form (\ref{adjunction1}),
the essence of ($\bigstar$) can be generalized to wider contexts as selection
criteria to choose states of relevance \cite{Ojim2002}.) From the conceptual
viewpoint, what is important here is that \textit{two different levels},
quatum statistical mechanics with family $K$ of mixtures of KMS states and
macroscopic thermodynamics described by $Th$ of probability measures of
fluctuating thermal parameters on the parameter space $B_{K}$, are so
interrelated by the two channels,\textit{\textbf{\ }c }$\rightarrow$\textit{q}
($\mathcal{C}^{\ast}$) and \textit{q }$\rightarrow$\textit{c }($(\mathcal{C}%
^{\ast})^{-1}$), that the following two points are simultaneously attained:

a) characterization of thermal reference states $K$ as image of $\mathcal{C}%
^{\ast}$, $\omega_{\rho}=\mathcal{C}^{\ast}(\rho)$: \textit{selection
criterion} for $K$,

b) \textit{thermal interpretation }of selected states in $K$ in terms of
classical data, $\Phi_{x}=\mathcal{C}(\hat{\Phi}(x))$ and $\rho=(\mathcal{C}%
^{\ast})^{-1}(\omega_{\rho})$.

Then the problem is now boiled down into how to select suitable classes of
\textit{non-equilibrium states }$\omega\notin K$ in such a way that some
thermal interpretations are still guaranteed. This is what to be answered in
the next section.

\section{Characterization of non-equilibrium local states by hierarchized
zeroth law of local thermodynamics and their thermal interpretations}

\textit{Hierarchized zeroth law of local thermodynamics}\textbf{\ }%
\cite{Oji02}: to meet simultaneously the two requirements of
\textit{characterizing an unkown state }$\omega$\textit{\ as a non-equilibrium
local state} and of \textit{establishing its thermal interpretation} in a
similar way to the above a) and b), we compare $\omega$ with thermal reference
states $\in K=\mathcal{C}^{\ast}(Th)$ by means of local thermal observables
$\in\mathcal{T}_{x}$ at $x$ whose physical meanings are exhibited by the
associated thermal functions\thinspace$\in\mathcal{C}(\mathcal{T}_{x})$.

In view of the above conclusion [\textit{q}$\rightarrow$\textit{c} channel
$(\mathcal{C}^{\ast})^{-1}$= thermal interpretation of quantum states] and
also of the hierarchy in $\mathcal{T}_{x}$, we relax the requirement for
$\omega$ to agree with $\exists\omega_{\rho}:=\mathcal{C}^{\ast}(\rho_{x})\in
K$ up to some suitable \textit{sub}space $\mathcal{S}_{x}$ of $\mathcal{T}%
_{x}$. Then, we characterize $\omega$ as a non-equilibrium local state by the
equalities
\begin{equation}
\omega(\hat{\Phi}(x))=\omega_{\rho_{x}}(\hat{\Phi}(x))=\mathcal{C}^{\ast}%
(\rho_{x})(\hat{\Phi}(x))
\end{equation}
valid for $\forall\hat{\Phi}(x)\in\mathcal{S}_{x}$. Namely, the unknown
$\omega$ should \textit{look like} a thermal reference state\textit{\ }%
$\omega_{\rho_{x}}$ \textit{as far as }the thermal properties described by
$\hat{\Phi}(x)\in\mathcal{S}_{x}$ are concerned. We denote this selection
criterion by
\begin{equation}
\omega\underset{\mathcal{S}_{x}}{\equiv}\mathcal{C}^{\ast}(\rho_{x}),
\label{S-thermal}%
\end{equation}
and call such $\omega$ an $\mathcal{S}_{x}$\textit{-thermal }state. In terms
of thermal functions $\Phi_{x}:=\mathcal{C}(\hat{\Phi}(x))\in\mathcal{C}%
(\mathcal{S}_{x})$, this can be rewritten as
\begin{equation}
\omega(\Phi)(x):=\omega(\hat{\Phi}(x))=\rho_{x}(\Phi_{x}).
\end{equation}
So, $\omega$: $\mathcal{S}_{x}$\textit{-thermal }implies that the selection
criterion $\omega\underset{\mathcal{S}_{x}}{\equiv}\mathcal{C}^{\ast}(\rho
_{x})$ can be \textquotedblleft solved\textquotedblright\ conditionally in
favour of $\rho_{x}$ as $``(\mathcal{C}^{\ast})^{-1}"(\omega)\underset
{\mathcal{C}(\mathcal{S}_{x})}{\equiv}\rho_{x}$, which provides the local
thermal interpretation of $\omega$ \cite{Oji02}. Physically this means the
state $\omega$ looks like a thermal equilibirum $\mathcal{C}^{\ast}(\rho_{x})$
\textit{locally} at $x$ to within a level controlled by a subset
$\mathcal{S}_{x}$ of thermal observables.

To be precise mathematically, we need here to be careful about the meaning of
such a heuristic expression as $``(\mathcal{C}^{\ast})^{-1}"(\omega)$ for
$\omega\notin K$ in relation to our observation above: $\omega\notin
K=\mathcal{C}^{\ast}(Th)$. As we shall see below, $``(\mathcal{C}^{\ast}%
)^{-1}"$ outside of $K$ is certainly \textit{not} a \textit{q}$\rightarrow
$\textit{c} channel preserving the positivity, whereas it can be seen to be
still definable on the states $\omega$ selected out by the above criterion
Eq.(\ref{S-thermal}), by means of its equivalent reformulation given by:

\begin{criterion}
For a subspace $\mathcal{S}_{x}$ of $\mathcal{T}_{x}$ containing $\mathbf{1}$,
a state $\omega\in E_{x}$ is $\mathcal{S}_{x}$\textit{-thermal} iff there is a
compact set ${B}\subset V_{+}$ s.t.
\begin{align}
|\omega(\hat{\Phi}(x))|  &  \leq\tau_{B}(\hat{\Phi}(x)):=\sup_{(\beta,\mu)\in
B_{K},\beta\in B}\,|\omega_{\beta,\mu}(\hat{\Phi}(x))|\nonumber\\
&  =\left\vert \left\vert \mathcal{C}(\hat{\Phi}(x))\right\vert \right\vert
_{B},\quad\text{for }\hat{\Phi}(x)\in\mathcal{S}_{x}. \label{norm}%
\end{align}

\end{criterion}

\noindent(The above semi-norm is well-defined under the condition that
$B_{K}\ni(\beta,\mu)\longmapsto\omega_{\beta,\mu}\in K$ is (weakly)
continuous, which requires singularities of critical points to be excluded
from our considerations.)

While the requirement for $``(\mathcal{C}^{\ast})^{-1}"(\omega)$ to be a
probability measure forces $\omega$ to be among the reference states belonging
to $K$, the above inequality (\ref{norm}) combined with the Hahn-Banach
extension theorem (under the assumption for $\tau_{B}$ to be a norm) allows us
to extend $\mathcal{C}(\mathcal{S}_{x})\ni\mathcal{C}(\hat{\Phi}%
(x))\longmapsto\omega(\hat{\Phi}(x))$ as a \textit{linear functional} defined
on $\mathcal{C}(\mathcal{S}_{x})$ to one $\nu$ defined on $\overline
{\mathcal{C}(\mathcal{T}_{x})}=C(B_{K})$, which should \textit{not} be a
positive-definite measure but is a \textit{signed }measure: $\nu=\nu_{+}%
-\nu_{-}$, $0\leq\nu_{\pm}\in C(B_{K})_{+}^{\ast}$, $\nu_{-}\neq0$, $\nu
_{-}\upharpoonright_{\mathcal{C}(\mathcal{S}_{x})}=0$, $\mathcal{C}^{\ast}%
(\nu_{+})\upharpoonright_{\mathcal{S}_{x}}=\omega$ $\upharpoonright
_{\mathcal{S}_{x}}$. The similar argument for this has already been used in
\cite{BOR} to ensure the existence of a genuine \textit{non-equilibrium} local
state $\omega$ which is $\mathcal{S}_{x}$-thermal with finite-dimensional
subspace $\mathcal{S}_{x}$ at a lower level of hierarchy in $\mathcal{T}_{x}$,
but which shows deviations from $K$ for observables outside of $\mathcal{S}%
_{x}$. (See this discussion also for the case with $\tau_{B}$ being a
\textit{semi-}norm.) Thus, understanding the meaning of $(\mathcal{C}^{\ast
})^{-1}(\omega)$ as the set of inverse images of $\omega$ under $\mathcal{C}%
^{\ast}$ in the space $C(B_{K})^{\ast}$ of linear functionals,
\begin{align}
(\mathcal{C}^{\ast})^{-1}(\omega):=  &  \{\nu\in C(B_{K})^{\ast};\nu=\nu
_{+}-\nu_{-},\nu_{\pm}\geq0,\nonumber\\
&  \nu_{-}\upharpoonright_{\mathcal{C}(\mathcal{S}_{x})}=0,\mathcal{C}^{\ast
}(\nu_{+})\upharpoonright_{\mathcal{S}_{x}}=\omega\upharpoonright
_{\mathcal{S}_{x}}\};
\end{align}
we can put Eq.(\ref{S-thermal}) into the similar form to Eq.(\ref{adjunction1}%
) as%
\begin{equation}
E_{x}(\omega,\mathcal{C}^{\ast}(\rho_{x}))/\mathcal{S}_{x}\overset
{q\rightleftarrows c}{\simeq}Th((\mathcal{C}^{\ast})^{-1}(\omega),[\rho
_{x}])/\mathcal{C}(\mathcal{S}_{x}), \label{adjunction2}%
\end{equation}
where $[\rho_{x}]:=\{\sigma\in Th;\sigma\upharpoonright_{\mathcal{C}%
(\mathcal{S}_{x})}=\rho_{x}\upharpoonright_{\mathcal{C}(\mathcal{S}_{x})}\}$
enters here owing to the non-uniqueness of $\rho_{x}\in Th$ in
Eq.(\ref{S-thermal}). This relation can be viewed as a form of
\textquotedblleft\textit{hierarchized zeroth law of local thermodynamics}%
\textquotedblright; the reason for mentioning the \textquotedblleft zeroth
law\textquotedblright\ here is due to the implicit relevance of measuring
processes of local thermal observables validating the above equalities, which
require the \textit{contacts of two bodies}, measured object(s) and measuring
device(s), in a local thermal equilibrium, conditional on the chosen
$\mathcal{S}_{x}$. (The transitivity of this contact relation just corresponds
to the localized and hierarchized version of the standard zeroth law of thermodynamics.)

It is interesting to note that, in view of the relation
\begin{align}
\exists\nu &  =\nu_{+}-\nu_{-}\in(\mathcal{C}^{\ast})^{-1}(\omega)\text{ with
}\nu_{-}=0\Longleftrightarrow(\mathcal{C}^{\ast})^{-1}(\omega)=\{\nu\}\subset
Th\nonumber\\
&  \Longleftrightarrow\omega\in K\Longleftrightarrow\text{[maximal choice of
}\mathcal{S}_{x}^{\prime}\text{ s.t. }\mathcal{C}^{\ast}(\nu_{+}%
)\upharpoonright_{\mathcal{S}_{x}^{\prime}}=\omega\upharpoonright
_{\mathcal{S}_{x}^{\prime}}\text{]}=\mathcal{T}_{x},
\end{align}
we can specify the extent to which a non-equilibrium $\mathcal{S}_{x}$-thermal
$\omega$ deviates from equilibria belonging to $K$ by the \textit{failure of
state positivity} ($\nu_{-}\neq0$) and can also measure it by the
\textit{maximal size} of $\mathcal{S}_{x}^{\prime}$ within the hierarchy of
subspaces $\mathcal{S}_{x}^{\prime}$ in $\mathcal{T}_{x}$ such that
$\mathcal{S}_{x}^{\prime}\supset\mathcal{S}_{x}$, $\nu_{-}\upharpoonright
_{\mathcal{C}(\mathcal{S}_{x}^{\prime})}=0$ with all the possible choices of
$\nu\in(\mathcal{C}^{\ast})^{-1}(\omega)$: owing to the presence of $\nu_{-}$,
$\omega$ ceases to be $\mathcal{S}_{x}^{\prime}$-thermal when $\mathcal{S}%
_{x}^{\prime}$ is so enlarged that $\nu_{-}\upharpoonright_{\mathcal{C}%
(\mathcal{S}_{x}^{\prime})}=0$ is invalidated, which shows that $\omega$
shares with reference states in $K$ only gross thermal properties described by
smaller $\mathcal{S}_{x}^{\prime}$. In this sense, the hierarchy of
$\mathcal{S}_{x}^{\prime}$ in $\mathcal{T}_{x}$ should have a close
relationship with the thermodynamic hierarchy at various scales appearing in
the transitions between non-equilibrium and equilibrium controlled by certain
family of \textit{coarse graining} procedures. Thus, we see that our selection
criterion can give a characterization of states identifiable as
non-equilibrium ones and, at the same time, provide associated relevant
physical interpretations of the selected states in a systematic way.

\noindent\newline---Fluctuations of thermal quantities; temperature as a
physical quantity---

The present framework allows one also to judge whether a thermal function
$\Phi$ has locally a sharply specified value in a state $\omega$ or is
statistically fluctuating, which can be implemented if $\mathcal{S}_{x}$ is
large enough for the \textit{mean value} of $\Phi_{x}$ together with its
\textit{fluctuations} to be determined within it. For instance, if
$\mathcal{S}_{x}$ contains local observables $\hat{\Phi}_{1}(x)$ and
$\hat{\Phi}_{2}(x)$ corresponding respectively to $\Phi_{x}=\mathcal{C}%
(\hat{\Phi}_{1}(x))$ and $\Phi_{x}^{2}=\mathcal{C}(\hat{\Phi}_{2}(x))$, we
have a thermal function $(\Phi_{x}-\kappa\,1)^{2}=\mathcal{C}(\delta\hat{\Phi
}_{\kappa}(x))$ with
\begin{equation}
\delta\hat{\Phi}_{\kappa}(x):=\hat{\Phi}_{2}(x)-2\kappa\,\hat{\Phi}%
_{1}(x)+\kappa^{2}\,\mathbf{1},\quad\kappa\in\mathbb{R}.
\end{equation}
Since this is non-negative in all thermal reference states $\in K$ and
vanishes only in those states with $\Phi_{x}$ having a sharp value $\kappa$,
we can conclude that $\Phi_{x}$ at $x$ has the sharp value $\kappa$ in such an
$\mathcal{S}_{x}$-thermal state $\omega$ that $\omega(\delta\hat{\Phi}%
_{\kappa}(x))=0$.

In virtue of this scheme it is meaningful to treat the (inverse) temperature
$\beta$ as a real physical quantity to be determined a posteriori through its
measurements, which is in sharp contrast to the standard idea in statistical
mechanics of treating it as an a priori given \textit{parameter.} Choosing
suitable spaces $\mathcal{S}_{x}$ (which is finite dimensional in generic
cases \cite{BOR}), we can select a state $\omega$ having locally a
\textit{sharp temperature} vector $\beta_{x}$, i.e., $\omega\underset
{\mathcal{S}_{x}}{\equiv}\omega_{\beta_{x}}$. In this case, all thermal
functions $\Phi_{x}$ corresponding to $\hat{\Phi}(x)\in\mathcal{S}_{x}$ have
locally definite values. Concerning the possible objections against the
introduction of a temperature $\beta_{x}$ (as well as any kind of thermal
objects) \textit{at a point} $x$ (without any extension), it is important to
recall here that any state $\omega\in E_{x}$ relevant to our present context
is already \textit{extended in spacetime} effectively, owing to the regularity
condition (\ref{regularity}) imposed on it. In this sense, the point $x$ in
$\beta_{x}$ should not literally be understood to refer to a strictly
microscopic spacetime point but to a macroscopic one with certain fuzzy
extensions, which is taken into account on the side of chosen states.

\noindent\newline---Space-time evolution of thermal properties---

Extending our formalism from a point $x$ to a \textit{(finitely extended)
subregion} $\mathcal{O}\subset\mathbb{R}^{4}$, we can now incorporate
\textit{local states with thermal interpretation in }$\mathcal{O}$. For
simplicity, we keep the set of thermal functions fixed in each region, by
identifying the spaces $\mathcal{S}_{x}$, $x\in\mathcal{O}$ through
translations:
\begin{equation}
\mathcal{S}_{x}:=\alpha_{x}(\mathcal{S}_{0}),\quad x\in\mathcal{O}.
\end{equation}
With this convention understood, we say that a state $\omega\in E_{\mathcal{O}%
}$ is $\mathcal{S}_{\mathcal{O}}$\textit{-thermal in }$\mathcal{O}$, if there
exists $\omega_{\rho(x)}\in K$ for each $x\in\mathcal{O}$, s.t. $\omega
\underset{\mathcal{S}_{x}}{\equiv}\omega_{\rho(x)}$. The resulting functions
$\mathcal{O}\ni x\mapsto\omega(\Phi)(x):=\omega(\hat{\Phi}(x))$ describe the
space-time behaviour of mean values of thermal functions $\Phi$. Hence they
provide the \textit{link between microscopic dynamics }$\alpha_{x}$
\textit{and the evolution } $x\mapsto\omega(\Phi)(x)$ \textit{of macroscopic
thermal properties}, i.e. \textit{thermo-dynamics} of states. We have reached
the same level as the familiar local formulation of manifolds at the beginning.

On this setting, thermal functions in a state near equilibrium are generally
shown to satisfy a linear evolution equations, which can be viewed as a
generalization of low energy theorems to thermal situations. This is
consistent with interpretation of perturbations to equilibrium in terms of quasi-particles.

The two goals of identifying non-equilibrium local states admitting local
thermal interpretation and of describing their specific thermodynamic
properties are solved simultaneously by the above selection criterion based
upon a \textit{localized and hierarchized form of the zeroth law} of
thermodynamics. In this framework, we have identified at least three different
kinds of sources of derivations of an $\mathcal{S}_{x}$-thermal
non-equilibrium local state $\omega\in E_{x}$ from the genuine equilibrium
states $\omega_{\beta}$ as 

\begin{itemize}
\item[a)] \textit{spacetime dependence} of thermal parameters such as
temperature distributions $x\longmapsto\beta(x)$, 

\item[b)] \textit{statistical fluctuations} of thermal parameters at $x$
described by probability distributions $d\rho_{x}(\beta)\in Th$, 
\end{itemize}
\noindent and 
\begin{itemize}
\item[c)] essential deviations of local states $\omega\in E_{x}$ from states
in $K$ expressed by the \textit{positivity-violating} term $\nu_{-}\neq0$ in
$\nu=\nu_{+}-\nu_{-}\in(\mathcal{C}^{\ast})^{-1}(\omega)\subset C(B_{K}%
)^{\ast}$ with $\nu_{-}\upharpoonright_{\mathcal{C}(\mathcal{S}_{x}%
)}=0,\mathcal{C}^{\ast}(\nu_{+})\upharpoonright_{\mathcal{S}_{x}}%
=\omega\upharpoonright_{\mathcal{S}_{x}}$. 
\end{itemize}

\noindent For concrete examples to exhibit the basic features, see \cite{BOR}.
Here discussions have been focused on the conceptual aspects developed in
\cite{Oji02}.

\section{General-relativistic extension: global vs. local and flat vs. curved}

\noindent--Global vs. local--

Looking back over what is done so far, we notice here some room for further
improvements and generalizations in view of such restricted choices of
\textit{global} KMS states \cite{OK} on the side of thermal reference states.
When compared with our motivating discussion of manifolds, it also looks
strange that the whole theory is restricted only to within the \textit{flat}
Minkowski spacetime \cite{OF} in spite of the emphasis on the \textit{local
}aspects of states to be examined. Actually, in the comparison of an unknown
state $\omega$ with a known reference state $\omega_{\rho}\in K$, the latter
serves only to provide a reference data set $\omega_{\rho}(A)$ involving
\textit{local} thermal observables $A\in\mathcal{T}_{x}$ or $\mathcal{S}_{x}$
near the focus point $x$, which requires only the \textit{local restrictions}
of states $\omega_{\rho}\in K$ onto small neighbourhoods of $x$ (i.e., local
state germs in $K\hookrightarrow E_{x}$). Therefore, once the reference states
$\omega_{\rho}$ can properly be specified \textit{within small neighbourhoods}
of spacetime, we expect the freedom to go across the barriers separating the
flat and curved spacetimes, existing at the \textit{global} level but
irrelevant locally. Such a possibility can naturally be read off in our
selection criterion formulated as a relation of \textit{comparison} (more
appropriately, a categorical adjunction), Eq.(\ref{adjunction2}), which is
\textit{not} necessarily required to be a strict equality between $\omega$ and
$\mathcal{C}^{\ast}(\rho)$ but should be a well-defined and suitably
controllable \textit{relation}, as is common in many cases of reference to
standard objects constituting a \textit{model space}, such as local charts
referring to $\mathbb{R}^{n}$ in manifolds. On the basis of this observation,
a more flexible setting up is envisaged to emerge through such a possibility
that unknown $\omega$ to be examined can be generalized to those states living
in \textit{curved} background spacetimes: the mathematical basis for this
physical ideas can already be found in the notions of local definiteness
and/or local normality (see \cite{Haag, HNS}), which allow one to treat
generic localized states of quantum fields in curved spacetime backgrounds in
the same Hilbert space of the vacuum representation in the Minkowski spacetime.

\noindent\newline--Flat vs. curved spacetimes--

From the above point of view, we try now to extend the previous scheme to the
situation with quantum fields in a \textit{curved} spacetime by restricting
the original thermal reference states to \textit{local\textbf{\ }}small
regions in the \textit{flat} Minkowski spacetime. Note here that, in sharp
contrast to the examined unknown $\omega$ being allowed to be a state in a
\textit{curved} spacetime, the reference states $\omega_{\rho}$ are understood
as the local restrictions of mixtures of global KMS states still living in the
\textit{flat} Minkowski spacetime; this is not only in harmony with the line
of thought found in our starting discussion of manifolds taking the flat
$\mathbb{R}^{n}$ as the model space, but also is a very important and
inevitable choice necessitated by the possible \textit{absence of appropriate
vacua and/or KMS states }in generic curved spacetimes.

In spite of the \textit{inherent slight delocalization }of our selected states
$\omega$ due to the condition (\ref{regularity}), we can here benefit from the
expressions referring to one spacetime point $x$\textit{\textbf{\ }}as
follows. To a small neighbourhood of a point $x$ in a curved spacetime $M$ we
can apply Einstein's basic idea of \textit{equivalence principle }based upon
the \textit{free-falling frame} at $x$ whose mathematical expression can be
found in the notion of \textit{normal coordinates }\cite{KobNom1}, the
coordinates along geodesic flows starting from $x$\textit{\ }which are always
definable in some neighbourhood $\mathcal{O}_{0}$ of the origin $0$ of the
tangent space $T_{x}(M)$ at $x$ even for \textit{incomplete} geodesic flows,
$Exp_{x}:(T_{x}(M)\supset)\mathcal{O}_{0}\rightarrow Exp_{x}(\mathcal{O}%
_{0})=\mathcal{O}_{x}(\subset)M$. Corresponding to this local diffeomorphism
$Exp_{x}$, we can define a mapping $\Phi_{x}$ to transform locally a QFT
defined in a flat $\mathcal{O}_{0}(\hookrightarrow\mathbb{R}^{4})$ into the
local restriction of a QFT in curved $M$ on its small neighbourhood
$\mathcal{O}_{x}$ of $x$ (owing to the \textit{functoriality} in the
definition of quantum fields on curved spacetimes \cite{Oji86, BFV}):
$\Phi_{x}:\mathcal{A}_{0}(\mathcal{O}_{0})\rightarrow\mathcal{A}%
_{M}(\mathcal{O}_{x})$, where $\mathcal{O}\longmapsto\mathcal{A}%
_{0}(\mathcal{O)}$ and $\mathcal{O}\longmapsto\mathcal{A}_{M}(\mathcal{O)}$
denote the corresponding local nets, respectively, on the flat Minkowski
spacetime and on a curved spacetime $M$. Then what we need is simply to modify
our selection criterion Eq.(\ref{adjunction2}) into $E_{M,x}(\omega
,\mathcal{C}_{x}^{\ast}(\rho))/\Phi_{x}(\mathcal{S}_{0}):=E_{0}(\omega
\circ\Phi_{x},\mathcal{C}^{\ast}(\rho))/\mathcal{S}_{0}\overset
{q\rightleftarrows c}{\simeq}Th((\mathcal{C}_{x}^{\ast})^{-1}(\omega
),[\rho])/\mathcal{C}(\mathcal{S}_{0})$, where $E_{M,x}$ and $\Phi
_{x}(\mathcal{S}_{0})$ are the sets of local thermal states and of local
thermal observables, respectively, at $x$ in the curved spacetime $M$ and
$\mathcal{C}_{x}=\mathcal{C}\circ\Phi_{x}^{-1}$. In this way, our formulation
of non-equilibrium local states can safely be extended to the
general-relativistic context by simple restriction of reference states onto
small neighbourhoods. (While the set $K_{M,x}$ of reference states at $x$ in a
curved spacetime $M$ can simply be defined here as the image $(\Phi_{x}^{\ast
})^{-1}(K)$ of the corresponding set $K$ on the flat spacetime through the
normal coordinates $Exp_{x}$ in the $C^{\infty}$-context, the intrinsic
characterization of the notions in $M$ related to the KMS condition will
certainly require the analytic structure on $M$, as is the case in the
discussion of analytic wavefront sets in \cite{SVW}.)

\noindent\newline--Interacting vs. free--

In view of our focus on small neighbourhoods of a spacetime point, the local
normality allows us to treat the effects of \textit{interactions} as a
perturbation to the \textit{free} dynamics, without complications related to
the thermodynamic limit such as the Haag theorem, at least, at the abstract
levels. Therefore, it will be very convenient if we can choose \textit{free
field} models for the reference system \cite{OK}; this will not only make the
reference states more accessible to the practical computations, but also
conceptually appealing in relation to the physical origin of temperature
referring to the \textit{ideal gas} in the equation of states as well as
Bolzmann's kinetic definition of it. To implement this idea, however, we
should solve the difficulty taking the familiar form of \textit{ultraviolet
divergences}. While the combined use of the regularity condition
(\ref{regularity}) imposed on the states of relevance and the normal products
in OPE can remove divergences order by order systematically in the
perturbation, the available methods are not powerful enough to remove the
full-order divergences at once, in spite of all the up-dated attractive tools
for regularizing and renormalizing these divergences (e.g., the use of energy
bounds, OPE \cite{Bo}, local perturbation scheme \cite{BFK} of Epstein-Glaser
type and Connes-Kreimer method \cite{CK} of renormalization). If we succeed in
finding a satisfactory reason for terminating the perturbative expansions at
some finite orders, then the algebraically formulated \textit{local}
perturbation scheme \cite{BFK} will become physically relevant. Aside from
this long-standing problem, what is also important in the present context is
to attain the effective separation between \textit{order parameters} to
describe non-trivial structures in phase diagrams due to interactions and
small \textit{fluctuations} within fixed phases which are expected to be
described by particle-like modes (\cite{BrBuch}). In any case, it seems still
premature to expect a practical and satisfactory solution to our desire in
this direction.

Extrapolating the above logical lines, we can formulate \cite{Ojim2002} a
unified scheme for generalized sectors (discrete and/or continuous) based upon
selection criteria, which provides new physical operational interpretations of
superselection theory, extends it to spontaneously broken symmetries, and
exhibits close relationship of basic notions in quantum measurement theory
with those in control theory \cite{ArbibMane}, such as the notions of
\textit{measurement scheme} \cite{Ozawa}\ and \textit{state preparation}
processes in connection with \textit{realizability} and \textit{reachability},
respectively, constituting the core of the latter. It would be important to
note that \textit{applicability domain} of a given theory is, in principle,
encoded in this scheme in the \textit{matching} relation between a chosen
selection criterion for relevant states and the available observables.\newline%
\newline

\end{document}